\documentclass[prl,aps,twocolumn,groupedaddress,showpacs,floatfix]{revtex4}
\usepackage{amsmath,amssymb,multirow,epsfig,bm,subfigure}

\newcommand{\beq}{\begin{equation}}
\newcommand{\eeq}{\end{equation}}
\newcommand{\bea}{\begin{eqnarray}}
\newcommand{\eea}{\end{eqnarray}}

\begin{document}
\title{Signatures of a gap in the conductivity of graphene}

\author{Joaqu\'{\i}n E. Drut$^1$, Timo A. L\"ahde$^2$ and Eero T\"ol\"o$^2$}
\affiliation{$^1$Department of Physics, The Ohio State University, Columbus, OH 43210--1117, USA}
\affiliation{$^2$Helsinki Institute of Physics and Department of Applied Physics,
Aalto University, FI-02150 Espoo, Finland}

\begin{abstract}
The electrical conductivity of suspended graphene has recently been measured for the first time,
and found to behave as $\sigma\sim\sqrt{|n|}$ as expected for Dirac quasiparticles at large carrier density.
The charge inhomogeneity is strongly reduced in suspended samples, which revealed an unexpected
insulating trend in $\sigma(T)$. Above a transitional density $n^*_{}$, the temperature dependence was found 
to revert to metallic. We show that these features of the DC~conductivity are
consistent with a simple model of gapped Dirac quasiparticles, with specific signatures of a gap in the vicinity
of the charge-neutrality point. Our findings are reminiscent of the conductivity profile in semiconducting 
materials, exhibiting a thermal activation for $T \geq \tilde T$ and a weakly $T$-dependent
background for $T \leq \tilde T$, where $\tilde T$ is given by the saturation density $\tilde n$ associated
with the residual charge inhomogeneity. We discuss possible origins of a bandgap in graphene as well
as alternative scenarios.

\end{abstract}

\date{\today}

\maketitle

The experimental study of graphene has recently progressed towards an improved understanding of the 
intrinsic properties of this new carbon nanomaterial. Notably, a clear signal for the fractional quantum 
Hall effect has been found~\cite{FQHE_Andrei, FQHE_Kim}, and carrier mobilities far in excess of 
silicon-based devices have been reached~\cite{Bolotin1}. While the high quality of graphene and the Dirac 
nature of the charge carriers is by now well established, the issue of gap formation has remained far less 
clear. Indeed, several features suggestive of a gap but difficult to interpret have been reported under 
various circumstances. The Lanzara group has performed ARPES studies~\cite{Lanzara1, Lanzara3} of graphene 
layers on a substrate, whereby a gap-like feature was detected and attributed to substrate effects. On the 
other hand, the conductivity of suspended graphene devices was characterized by the Kim 
group~\cite{Bolotin2} and the Andrei group~\cite{Andrei_SG}. In addition to demonstrating a strongly 
$T$-dependent DC conductivity $\sigma$, a transitional carrier density $n^*_{}$ was found, where $\sigma(T)$ 
changes from metallic to insulating. A different perspective is provided by the Andrei group in 
Ref.~\cite{Andrei_GG} via STM spectroscopy of decoupled graphene flakes on graphite, where a $\sim 10$~meV 
gap centered on the Dirac point was found at zero magnetic field, accompanied by a corresponding splitting 
of the lowest Landau level.

On the theoretical side, a quantitative explanation of $\sigma(n,T)$ has remained elusive. For instance, the 
appearance of a transitional density $n^*_{}$ is incompatible with electron-phonon 
scattering~\cite{Bolotin2, phonons}, while electron-electron scattering leads to metallic behavior at the 
neutral point~\cite{ee_int}. A qualitative description of the empirical $\sigma(n,T)$ is given by Landauer 
transport theory for ballistic graphene~\cite{FabryPerot}. The Hall probe lead geometry of 
Ref.~\cite{Bolotin2} minimizes the effects of a finite sample size, whereas the two-lead geometry of 
Ref.~\cite{Andrei_SG} is better described in the Landauer approach. Our primary objectives are to determine
whether the measured $\sigma(T)$ of Ref.~\cite{Bolotin2} at the neutral point can be quantitatively described
in terms of free gapped Dirac quasiparticles and a weakly $T$-dependent background, and whether such
a description is consistent with $\sigma(n,T)$ at finite carrier density.

The rationale for the gapped quasiparticle scenario is as follows: Initially, the suspended samples of 
Ref.~\cite{Bolotin2} showed an essentially $T$-independent DC~conductivity, which upon current-annealing 
acquired a pronounced $T$-dependence of insulating type. While the increase in resistivity at low $T$ 
remained modest (a factor of $\sim\!3$ in the range $5 - 150$~K), the data suggests 
the existence of two regimes, reminiscent of conventional semiconductors~\cite{MottBook}: a low-$T$ regime 
($5 - 35$~K) where $\sigma$ is dominated by a weakly $T$-dependent background $\sigma^{}_{bg}$, and a 
thermally activated regime ($35 - 150$~K) where $\sigma$ increases rapidly with $T$. Notably, several 
mechanisms for gap generation have been proposed, such as explicit breaking of the sublattice symmetry by 
mechanical strain~\cite{strain}, or spontaneous induction of a Mott insulating state via strong Coulomb 
interactions~\cite{CastroNetoPhysics}. It is plausible that $\sigma^{}_{bg}$ is due to residual charge density 
inhomogeneities~\cite{puddles} which, however, are reduced by an order of magnitude compared 
with samples on a substrate~\cite{Andrei_SG}. Regardless of the origin of $\sigma^{}_{bg}$, we find that 
it can be reliably subtracted.

\begin{figure}[b]
\epsfig{file=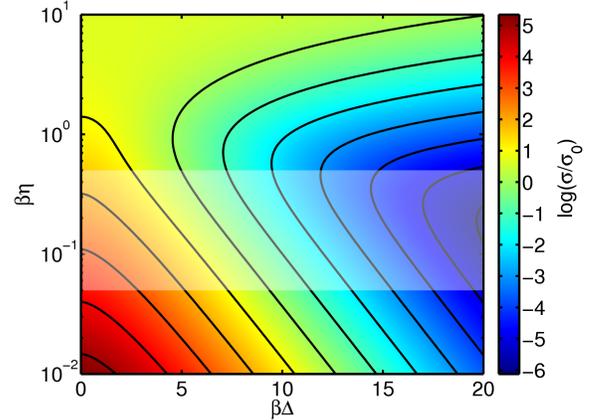, width=.9\columnwidth} 
\caption{(Color online) DC conductivity $\sigma_q^{}$ as a function of 
$\beta\Delta$ and $\beta\eta$, shown semilogarithmically. For $\beta\Delta \gg 1$, $\sigma_q^{}$ decreases
exponentially when $\beta\eta \ll 1$, while for $\beta\eta \gg 1$ any signature of a gap is washed out by the
large scattering rate. For $\beta\eta \ll 1$, $\sigma_q^{}$ decreases with $\beta\eta$, while for 
$\beta\eta \gg 1$ this behavior is reversed.
The shaded area denotes the approximate range in $\beta\eta$ for suspended graphene samples.
\label{fig:DC}}
\end{figure}

The Hamiltonian describing Dirac quasiparticles with a gap $\Delta$ and Fermi
velocity $v_F^{}\simeq c/300$ is given by
\bea
H &=& \sigma_1^{}v_F^{}k_1^{} + \sigma_2^{}v_F^{}k_2^{} + \sigma_3^{}\Delta/2,
\eea
where the $\sigma_i$ are Pauli matrices, and we account for the spin and valley degeneracy below.
From this starting point, $\sigma$ is calculated as the diagonal part of the conductivity 
tensor $\sigma_{\mu\nu}^{}$, given by the Kubo formula
\begin{eqnarray}
\sigma_{\mu\mu}^{} \!\!&=&\!\! \frac{\pi e^2}{\hbar} \!
\int_{-\infty}^{\infty} \!\! d\epsilon \, \mathrm{Tr} \left\{
[H,r_\mu^{}] \, \delta\left(H-\epsilon-\frac{\omega}{2}\right) [H,r_\mu^{}] \quad\quad \right. \\
&& \times \:\left. \delta\left(H-\epsilon+\frac{\omega}{2}\right) \right\}
\: \frac{f(\beta\epsilon+\frac{\beta\omega}{2}) 
- f(\beta\epsilon-\frac{\beta\omega}{2})}{\omega}, \nonumber
\end{eqnarray}
where $\beta \equiv 1/k_B^{}T$ and the Fermi function is given by
$f(x) = 1/(1+\exp(x))$. The contribution $\sigma_q^{}$ of the Dirac 
quasiparticles to the conductivity of a graphene monolayer is then
\begin{eqnarray}
\sigma_q^{} &=& \frac{4e^2}{h} \frac{\pi}{2}
\int_{-\infty}^{\infty} d\epsilon\, \int_{\Delta/2}^{\infty} d\xi\,\xi
\: \mathcal T_\omega^{}(\xi,\epsilon) \nonumber \\
&& \times \:\: \frac{f(\beta\epsilon-\frac{\beta\omega}{2}-\beta\mu) 
- f(\beta\epsilon+\frac{\beta\omega}{2}-\beta\mu)}{\omega},
\label{qp}
\end{eqnarray}
%
where $\mu$ is the chemical potential, and
the factor of~$4$ accounts for the spin and valley degrees of freedom. We find
\begin{eqnarray}
\label{Amplitude}
\mathcal T_\omega^{}(\xi,\epsilon) &\equiv& \frac{\xi^2+\Delta^2/4}{\xi^2} 
\left[
\delta_\eta^{}\left(\xi+\epsilon+\frac{\omega}{2}\right)
\delta_\eta^{}\left(\xi-\epsilon+\frac{\omega}{2}\right) \right. \nonumber \\
&& + \left.
\delta_\eta^{}\left(\xi+\epsilon-\frac{\omega}{2}\right)
\delta_\eta^{}\left(\xi-\epsilon-\frac{\omega}{2}\right) 
\right] 
\nonumber \\
&+& \frac{\xi^2-\Delta^2/4}{\xi^2}
\left[
\delta_\eta^{}\left(\xi-\epsilon-\frac{\omega}{2}\right)
\delta_\eta^{}\left(\xi-\epsilon+\frac{\omega}{2}\right) \right. \nonumber \\
&& + \left.
\delta_\eta^{}\left(\xi+\epsilon+\frac{\omega}{2}\right)
\delta_\eta^{}\left(\xi+\epsilon-\frac{\omega}{2}\right) 
\right], 
\end{eqnarray}
where $\eta$ is the scattering rate of the quasiparticles, which can be
accounted for~\cite{Ziegler} by broadening the delta functions according to $\pi\delta_\eta^{}(x) \equiv 
\eta/(x^2+\eta^2)$. While the $T$-dependence of $\eta$ is {\it a priori} unknown, we find that the
scenario of constant $\beta\eta$, which may be ascribed to scattering off impurities or thermally generated 
ripples~\cite{Ziegler}, is strongly favored by the available data in the range $35$~K~$ \leq T \leq 150$~K. 
The integral over $\xi$ in Eq.~(\ref{qp}) can be performed analytically, and in the DC limit it yields
\beq
\label{Amplitude_DC}
\int_{\Delta/2}^{\infty}\! d\xi\,\xi\,\mathcal T_0^{}(\xi,\epsilon) = 
\frac{1}{2\pi} - \frac{\Delta^2\!-4|z|^2}{16 \pi \epsilon \eta} \arg \left(\Delta^2 \!- 4 z^2\right), 
\eeq
where $z=\epsilon + i\eta$. The dependence of $\sigma_q^{}$ on $\beta\Delta$ and $\beta\eta$ is 
illustrated in Fig.~\ref{fig:DC}.

We now analyze the data of Ref.~\cite{Bolotin2} on the suspended graphene devices S1, S2 
and S3 in terms of the expression $\sigma \equiv \sigma_q^{} + \sigma_{bg}^{}$, where $\sigma_{bg}^{}$ 
denotes the background conductivity. The simplest form compatible with the low-$T$ data is the linear
one $\sigma_{bg}^{} \equiv \sigma_0^{}(1 - T_0^{}/T)$. A more specific choice is the
Variable-Range Hopping~(VRH) law $\sigma_{bg}^{} \equiv \sigma_0^{}\exp[- (T_0^{}/T)^\alpha]$
encountered in conventional semiconductors~\cite{MottBook}. While the former can be viewed as a
linearized VRH expression, it cannot remain valid at arbitrarily low~$T$ as it becomes unphysical in that limit.
As the $T$-dependence of the background is experimentally found to be weak, the choice of linear versus
VRH description has very little impact on the analysis.

\begin{figure}[t]
\epsfig{file=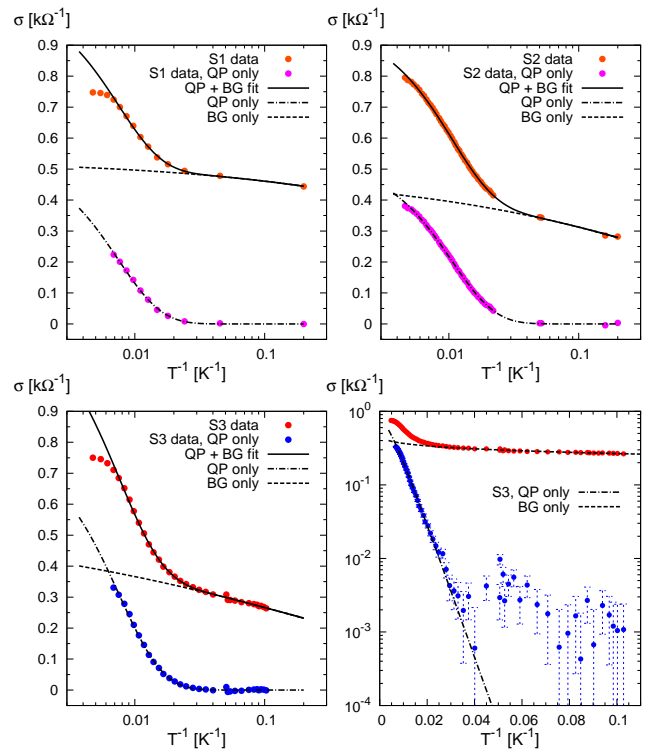, width=\columnwidth}
\caption{(Color online) Dirac Quasiparticle~(QP) and background~(BG) components (see Table~\ref{fit_table}) of the 
DC conductivity for the suspended graphene devices S1--S3, reproduced from Ref.~\cite{Bolotin2}. All devices 
show a ``knee" separating the thermally activated and background regions. After background subtraction, 
device S3 exhibits exponential behavior over more than two orders of magnitude. The slight curvature at 
high~$T$ is due to the finite scattering rate $\beta\eta$.
\label{fig:samples}}
\end{figure}

In order to determine $\sigma_{bg}^{}$ in an unbiased fashion, we first fix $\sigma_0^{}$ and $T_0^{}$ using 
the data in the low-$T$ region where thermal activation is negligible. The next step is to subtract 
$\sigma_{bg}^{}$ at all $T$, and determine $\beta\eta$ and $\Delta$ by fitting $\sigma_q^{}$ to the 
resulting dataset. Finally, the results were confirmed by a simultaneous fit of all four parameters to the 
full dataset. In all cases negligible variation was observed, indicating that $\sigma_0^{}$ and $T_0^{}$ are 
effectively uncorrelated with $\beta\eta$ and $\Delta$. The optimal parameter values are summarized in 
Table~\ref{fit_table}, and the results are plotted against the data of Ref.~\cite{Bolotin2} in 
Fig.~\ref{fig:samples}.

Our findings indicate that the suspended graphene devices of Ref.~\cite{Bolotin2} exhibit, upon subtraction 
of $\sigma_{bg}^{}$, a thermally activated component which can be well described in terms of Eq.~(\ref{qp}) 
from $T \sim 150$~K down to $T \sim 35$~K, where the signal is lost due to limited measurement accuracy. As 
shown in Fig.~\ref{fig:samples}, this corresponds to exponential behavior over more than two orders of 
magnitude, with bandgaps in the range $\sim 25 - 40$~meV. The value $\beta \eta \simeq 0.1$ obtained for 
samples S1 and S2 is consistent with the high carrier mobilities and long mean free paths reported in 
Ref.~\cite{Bolotin2}. Specifically, for $T=35 - 150$~K we find $\eta=3.5 - 15$~K, with corresponding mean 
free paths of $\hbar v_F^{}/\eta \sim 2.0 - 0.5~\mu$m. For such low values of $\beta\eta$, the rate of 
exponential decay is characterized by $\Delta$ whereas $\beta\eta$ mainly determines the amplitude. It is 
noteworthy that the values of $\beta\eta$ and $\Delta$ are roughly sample-independent. We have checked that 
fits with zero gap~($\Delta = 0$), constant $\eta$ or zero background are incompatible with the data. Above 
$T \sim 150$~K, the data deviate from a description with constant $\beta\eta$, which may be 
ascribed to increasing phonon scattering at high~$T$~\cite{Bolotin2}.

\begin{figure}[t]
\epsfig{file=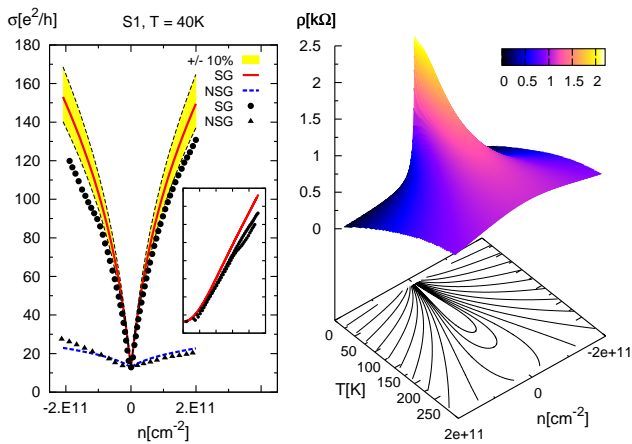, width=\columnwidth} 
\caption{(Color online) Left panel: Carrier density dependence of the
DC conductivity $\sigma \equiv \sigma_q^{} + \sigma_{bg}^{}$ 
for the suspended graphene device~S1 at $T=40$~K for our model (solid red line)
and experiment~\cite{Bolotin2} before (triangles) and after (circles) current-annealing. The dashed blue
line corresponds to $\beta\eta \sim 1.4$, although the properties before annealing are likely dominated by 
charged impurities~\cite{Bolotin2}.
Inset: $\sigma$ as a function of $\sqrt{|n|}$. Right panel: Resistivity $\rho(n,T)$ for device~S1.
As in Ref.~\cite{Bolotin2}, $\partial\rho/\partial T$ is metallic above $n^*_{} \sim 10^{10}_{}$ cm$^{-2}_{}$.
\label{fig:mu}}
\end{figure}

Once the description of the zero-density data is fixed, it is possible to predict
$\sigma$ at \emph{finite} carrier density $n$, with
\bea
n &\equiv& \frac{2}{\pi(\hbar v^{}_F\beta)^2_{}}
\int_0^\infty dx \,x \nonumber \\
&& \times \:
\left[g_{+}^{}(x,\beta \Delta,\beta \mu) - g_{-}^{}(x,\beta \Delta,\beta \mu)\right],
\eea
where $g_{\pm}^{}(x,\beta\Delta,\beta\mu) \equiv f(\sqrt{x^2 + (\beta \Delta/2)^2} \mp \beta\mu)$. It should
be emphasized that $\sigma(n)$ is strongly dependent on $\beta\eta$, which is obtained via analysis of
the quasiparticle contribution to $\sigma(T)$ at $n = 0$. 
In Fig.~\ref{fig:mu}, we show that the data of Ref.~\cite{Bolotin2} at finite $n$ is well described using the parameters of 
Table~\ref{fit_table}, as obtained from the gapped quasiparticle analysis at $n = 0$. Notice 
also that the experimentally observed $\sigma \sim \sqrt{|n|}$ dependence is reproduced (see 
Fig.~\ref{fig:mu}, inset), as well as the transitional carrier density $n^*_{} \sim 10^{10}$ cm$^{-2}$ which 
was identified in Ref.~\cite{Bolotin2} as separating metallic ($|n| > n^*_{}$) and insulating ($|n| < 
n^*_{}$) regimes in $\sigma(n,T)$. In the metallic regime, we recover the experimentally observed 
resistivity $\rho \sim T$.
	
\begin{table}[t]
\caption{Optimal parameter values for the suspended graphene devices of Ref.~\cite{Bolotin2}, corresponding
to the analysis in Fig.~\ref{fig:samples}. Similar results for the gap $\Delta$ and the scattering 
rate $\beta\eta$ were obtained by subtracting the background conductivity using a linear form or
a VRH description with $\alpha \sim 1/3$.
Entries labeled by an asterisk (*) indicate that $\beta\eta$ cannot be constrained
due to lack of information on the normalization factor $\sigma(5\text{K})$, which is not known for device~S3.
The resulting uncertainty in $\Delta$ is negligible.
\label{fit_table}}
\vspace{0.3cm}
\begin{tabular}{c||c|c|c|c|c}
sample & $\beta\eta$ & $\Delta$[meV] & $\sigma_0^{}$[k$\Omega^{-1}]$ & $T^{}_0$ [K] & $\chi^2/N_\text{dof}^{}$ 
\\ \hline\hline
S1($\alpha\!=\!1/3$)	& 	$0.103(3)$ & $36.8(1)$ & $0.530(6)$ & $ 0.027(7) $ & 1.6    \\
S2($\alpha\!=\!1/3$)	& 	$0.105(1)$ & $26.2(1)$ & $0.485(5) $ & $ 0.85(6) $ & 1.1    \\
S3($\alpha\!=\!1/3$)	& 	$0.070(1)^*$ & $35.8(1)$ & $0.471(2)$ & $ 1.87(8) $ & 4.1    \\
S3($\alpha\!=\!1/4$)	& 	$0.070(1)^*$ & $36.4(1)$ & $0.551(2)$ & $ 2.81(8) $ & 4.0    \\
S3(linear) & 			$0.070(1)^*$ & $32.0(1)$ & $0.339(1)$ & $ 2.21(1) $ & 7.1    \\
\end{tabular}
\end{table}

While a gap in the quasiparticle spectrum is an attractive interpretation of the thermally excited 
conductivity, it should be noted that a number of other mechanisms can yield similar results. These include 
localized disorder~\cite{Peres_disorder}, scattering by screened charged impurities~\cite{Hwang_DasSarma}, 
and transport gaps due to quantum confinement~\cite{FabryPerot}. Nevertheless, the gapped scenario is 
appealing as it provides a straightforward and consistent description of the $T$-~and $n$-dependence of 
$\sigma$, involving only quasiparticle excitations and an inhomogeneous regime close to the Dirac point. A 
plausible mechanism for gap generation in suspended graphene is given by the excitonic scenario, where a gap 
is dynamically generated by strong electron-electron interactions~\cite{Miranskyetal,Leal:2003sg}. This idea 
has recently been revived based on Lattice Monte Carlo simulations of the low-energy effective field theory 
of graphene~\cite{DrutLahde123,HandsStrouthos}. It is noteworthy that graphene flakes decoupled from 
underlying graphite layers have been found to exhibit a $\sim 10$~meV bandgap at zero magnetic field, with a 
corresponding splitting of the lowest Landau level of similar magnitude~\cite{Andrei_GG}.

What is the origin of the observed background conductivity? Empirically,
the physics at the Dirac point is obscured by charge density inhomogeneities~\cite{Bolotin2, Andrei_SG},
also referred to as ``puddles", where $n$ saturates to a finite value. The scale at which this
happens for presently available samples is $\tilde n \sim 10^{11}$~cm$^{-2}$ in non-suspended graphene,
and $\tilde n \sim 10^9$~cm$^{-2}$ in suspended graphene~\cite{Bolotin2,Andrei_SG}.
One can define energy and temperature scales $\tilde E$ and $\tilde T$ via
\bea
\tilde E &\equiv& k_B^{}\tilde T \simeq \hbar v_F^{} \sqrt{\pi\tilde n},
\eea
below which the description in terms of Dirac cones breaks down. Typically, $\tilde E \sim 40$~meV in 
graphene on a substrate, while in suspended samples $\tilde E < 10$~meV as $\tilde n$ is reduced by an order 
of magnitude, thereby making it possible to access the physics at the Dirac point. Thus a bandgap of $\Delta 
\sim 30$~meV is likely to be obscured by charge inhomogeneities in the non-suspended samples, whereas in suspended 
ones such a gap should be (partially) accessible. In samples with yet lower inhomogeneity, one should 
observe a decrease in the low-$T$ background conductivity and an enhancement of the strongly $T$-dependent 
quasiparticle contribution. A description of $\sigma_{bg}^{}$ which is compatible with data is given by the 
VRH model~\cite{MottBook}, which describes the residual conductivity in semiconducting materials when 
thermal excitation is negligible. If the background conductivity is indeed of the VRH form with 
$\alpha = 1/3$ (as expected in~2D systems), it may indicate hopping between localized states. 
However, the data are inconclusive as the variation of $\sigma$ is very mild at low~$T$.

In summary, we have explored the signatures of a bandgap in the DC~conductivity of graphene and showed that 
the empirical conductivity profiles $\sigma(n,T)$ are consistent with an interpretation in terms of Dirac 
quasiparticles with a non-zero bandgap $\Delta$. However, the associated thermally activated behavior is 
partially obscured below an empirically observed characteristic scale $\tilde n$, where the 
carrier density as a function of the gate voltage saturates. As gapped graphene is of great interest for 
nanoelectronic applications, a fully non-perturbative calculation (such as Lattice Monte Carlo) of the 
transport properties, including the effects of strong electron-electron interactions, is clearly called for. 
Such calculations also appear timely, as recent experimental work~\cite{Checkelsky,FQHE_Andrei,FQHE_Kim} at 
finite magnetic field has demonstrated a rich spectrum of phenomena closely related to Dirac physics at 
strong Coulomb coupling.

\begin{acknowledgments}

We acknowledge support under U.S. DOE Grants No.~DE-FG02-00ER41132 and DE-AC02-05CH11231, UNEDF SciDAC 
Collaboration Grant No.~DE-FC02-07ER41457 and NSF Grant No.~PHY--0653312. This study was supported in part 
by the Academy of Finland through its Centers of Excellence Program (2006 - 2011), the Vilho, Yrj\"o, and 
Kalle V\"ais\"al\"a Foundation of the Finnish Academy of Science and Letters, and by an allocation of 
computing time from the Ohio Supercomputer Center. We thank A.~H.~Castro Neto, R.~J.~Furnstahl, M.~Randeria, 
K.~I.~Bolotin, E.~Y.~Andrei, and P.~Hakonen for instructive discussions and comments. 
Part of this work used the {\tt CUBPACK} numerical quadrature routine.

\end{acknowledgments}



\end{document}